\documentclass[twocolumn,superscriptaddress,showpacs,prl]{revtex4}
\usepackage{graphicx}

\begin{document}
\draft
\title{Localized Character of $4f$ Electrons in CeRh$_x$(x=2,3) and CeNi$_x$(x=2,5)}

\author{Ran-Ju Jung}
\author{Byung-Hee Choi}
\author{Hyeong-Do Kim}
\author{S.-J. Oh}
\altaffiliation{Author to whom all correspondences should be
addressed. e-mail: sjoh@plaza.snu.ac.kr}
\affiliation{School of
Physics \& Center for Strongly Correlated Material Research, Seoul
National University, Seoul 151-742, Korea}

\author{En-Jin Cho}
\affiliation{Department of Physics, Chonnam National University,
Kwangju 500-757, Korea}

\author{T. Iwasaki}
\author{A. Sekiyama}
\author{S. Imada}
\author{S. Suga}
\affiliation{Department of Material Physics, Osaka University,
Osaka 560-8531, Japan}

\author{J.-G. Park}
\affiliation{Department of Physics, Sungkyunkwan University, Suwon
402-751, Korea}

\begin{abstract}
We have measured Ce $4f$ spectral weights of extremely
$\alpha$-like Ce-transition metal intermetallic compounds CeRh$_x$
(x=2,3) and CeNi$_x$ (x=2,5) by using the {\it bulk-sensitive}
resonant photoemission technique at the Ce
$M_5$($3d_{5/2}\rightarrow4f$)-edge. Unprecedentedly high energy
resolution and longer escape depth of photoemitted electron at
this photon energy enabled us to distinguish the sharp Kondo
resonance tails at the Fermi level, which can be well described by
the Gunnarsson-Sch\"onhammer(GS) calculation based on the Anderson
Impurity Hamiltonian.  On the other hand, the itinerant $4f$ band
description shows big discrepancies, which implies that Ce $4f$
electrons retain localized characters even in extremely
$\alpha$-like compounds.

\end{abstract}


\maketitle

Cerium is the first element of the $4f$ rare-earth series in the
periodic table, and occupies a special place in the condensed
matter physics in that its $4f$ electron is believed to lie on the
borderline between localization and itinerancy.  Its occupied $4f$
orbital is more extended than those of heavier rare earths, and it
is generally believed that an appropriate description of the
interaction between the $4f$ state and the conduction bands is
essential to understand the physics of cerium metal and
cerium-based compounds. The famous `$\gamma-\alpha$' phase
transition in cerium metal is a case in point.  This isostructural
transition is associated with a large volume change
($\approx15\%$) and loss of magnetism, and although it has been
investigated for more than 50 years the nature of phase transition
still remains controversial.  The early ``promotional
model''\cite{promotion} where one Ce $4f$ electron is presumed to
go into the $5d-6s$ conduction band in the $\alpha$ phase was not
supported by many experiments, and two other models based on quite
different starting points have emerged.

One is the Mott transition model advocated by B.
Johansson\cite{Johan}, which supposes that Ce $4f$ electron is
localized and non-bonding in the $\gamma$ phase, but becomes
itinerant and forms a $4f$ band in the small volume $\alpha$
phase.  The recent calculation on the ground state properties of
Ce metal based on the self-interaction corrected local (spin)
density approximation (SIC-L(S)DA) seems to support this
idea\cite{SIC}, and even the phase diagram has been calculated
\cite{PRLJOHAN}. The other model is the Kondo volume collapse
model\cite{allen,KVC}, which proposes that $4f$ electron is
localized in both $\gamma$ and $\alpha$ phases and the phase
transition is caused by the change in the conduction electron
screening of the $4f$ electron. In this model the Anderson
impurity hamiltonian is used to describe both spectroscopic and
thermodynamic properties\cite{Gunn,Baer}, and the
`$\gamma-\alpha$' phase transition was explained as due to the
variation of hybridization strength between $4f$ and conduction
states\cite{allen}.  Hence quite distinct pictures on the
electronic structure of $\alpha$-Ce have emerged from these two
models.

These contrasting views on the nature of Ce $4f$ electron, {\it
i.e.} ``localized'' vs ``itinerant'', extend to the understanding
of electronic structures of Ce compounds.  For example, when
cerium alloys with transition metal(TM) element to form
intermetallic compounds, the hybridization between the $4f$ state
and the $d$ state of TM can be much larger than the corresponding
hybridization in Ce metal\cite{Lars,Lukas}. In these so-called
extremely $\alpha$-like Ce-TM compounds, it was suggested that the
itinerant $4f$ picture forming a narrow band is rather proper, and
that this itinerant description makes correct predictions on the
equilibrium lattice constant and magnetic moment in agreement with
experiments for such compounds as CeFe$_2$, CeRh$_x$ (x=2,3), and
CeNi$_x$ (x=2,5)\cite{cetmb,ceni}. On the other hand, these
compounds have also been analyzed within the Anderson impurity
model\cite{alp}.

Photoelectron spectroscopy directly probes the electronic
structure (single-particle excitation spectrum in the many-body
description) of solids, and can in principle distinguish between
these two contrasting pictures. Indeed for CeRh$_3$, it was
claimed that photoemission and inverse photoemission spectra are
consistent with the $4f$ band picture\cite{Lukas,Kaindl}. But this
interpretation was challenged later\cite{Malterre}, while many
other Ce-TM compound photoemission data had been successfully
interpreted within the Anderson impurity model\cite{alp}. One
important factor contributing to this controversy is the fact that
most high-resolution photoemission experiments on Ce compounds so
far have been performed with low energy photons ($h\nu \leq$ 150
eV), which makes the spectra quite
surface-sensitive\cite{esc_depth}. Since the coordination number
becomes smaller at the surface, the hybridization of Ce 4$f$
states with TM $d$ levels is reduced, favoring localized $4f$
electron picture\cite{Duo}. Therefore, to understand the bulk
electronic structures it is necessary to separate out the surface
contribution from the bulk, but there are many uncertainties and
ambiguities in this procedure\cite{ceir2}. Hence it is desirable
to obtain {\it bulk-sensitive} $4f$ spectra from direct
experiments to resolve this controversy.

In this Letter, we report such study of the {\it bulk-sensitive}
$4f$ spectral weights in extremely $\alpha$-like Ce-TM
intermetallic compounds CeRh$_x$ (x=2,3) and CeNi$_x$ (x=2,5),
which are believed to be most likely to form itinerant $4f$ band
among Ce compounds.  Such bulk-sensitive high-resolution
photoemission experiments were made possible in a recently
developed synchrotron radiation beamline\cite{Saitoh}, where high
energy photons are incident to utilize the longer escape depth of
the emitted photoelectrons.  Since the photoionization cross
section of Ce $4f$ electron is usually much less than those of TM
$d$ electrons, we used the resonance photoemission (RPES)
technique at the Ce $3d$-edge to obtain the partial $4f$ spectral
weights, where the $4f$ emission is enhanced relative to other
conduction electron emissions by the process

\begin{displaymath}
 3d^{10}4f^1 +  \hbar \omega  \rightarrow  3d^94f^2 \rightarrow  3d^{10}4f^0 +
 photoelectron
\end{displaymath}

Similar resonance photoemission technique has been extensively
used at the Ce $4d$-edge to obtain $4f$ spectral weights of many
Ce compounds\cite{alp}.   The difference between RPES at these two
edges is that the kinetic energy of the valence photoelectron is
$\sim$ 880 eV at the Ce $3d$-edge, whereas it is $\sim$ 120 eV at
the $4d$-edge.  Hence the escape depth of photoemitted electron at
the $M_5$-edge is much longer than $4d$ RPES, and the spectra
becomes bulk sensitive with less than $\sim$ 15 \% of total
weights contributing from the surface region\cite{esc_depth}. This
technique has proven to be effective in elucidating bulk
electronic structures of several Ce compounds\cite{Sekiyama,
Iwasaki}.

All samples of CeRh$_2$, CeRh$_3$, CeNi$_2$ and CeNi$_5$ were
polycrystalline made by argon arc-melting followed by annealing,
and their crystal structures were checked by x-ray diffraction.
The Ce $3d\rightarrow4f$ RPES and x-ray absorption spectroscopy
(XAS) measurements at the Ce $M_{4,5}$ edge were performed in the
beamline BL25SU of SPring-8 in Japan. The energy resolution of the
photon source around Ce $M_{4,5}$ edge was better than 80 meV
(Full width at half maximum : FWHM) and the overall experimental
resolution $\sim$~100~meV FWHM was obtained by SCIENTA SES200
electron analyzer.  The pressure in the vacuum chamber was better
than $4\times10^{-10}$ Torr during the measurements. The data were
taken at 20 K and temperature was controlled by closed-cycle He
cryostat. Sample surfaces were cleaned by filing with a diamond
file {\it in situ} and we checked the cleanliness of the surface
by monitoring O 1$s$ level. The $E_{\rm F}$ of the sample was
referred to that of surface-cleaned Pd metal.

\begin{figure}
\includegraphics[width=0.35\textwidth]{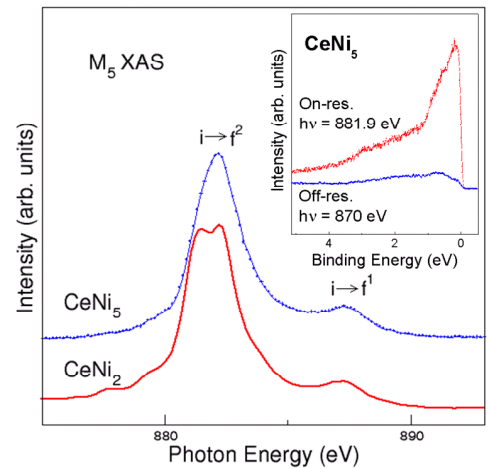}
\caption[] {The Ce $M_5$ XAS spectra of CeNi$_x$ ($x$ = 2, 5). The
inset shows off- and on-resonance photoemission data on CeNi$_5$.}
\label{fig:4}
\end{figure}

Figure 1 shows the x-ray absorption spectra of CeNi$_x$ (x= 2, 5)
at the Ce $M_5$ edge corresponding to the transition $3d_{5/2}
\rightarrow 4f$.  The lineshape of the main peak structure near
$h\nu$ = 882.3 eV is primarily determined by the multiplet
structures of the $ \underline{3d}f^2 $ electronic configuration,
where the underline represents a hole\cite{Jo88}. The slight
change of this lineshape and the satellite structure near
$\sim$885eV can be understood as the effect of hybridization
between Ce $4f$ electron and the valence band within Anderson
impurity model \cite{Fuggle83,jnj}. In the inset, we show the
photoemission spectra of CeNi$_5$ taken at photon energy below
this edge ($h\nu$ = 870 eV; off-resonance) and very close to the
maximum ($h\nu$ = 881.9 eV; on-resonance). We can see the drastic
change of spectral shapes due to the much enhanced Ce $4f$
emissions on resonance. We then extract the {\it bulk-sensitive}
$4f$ spectral weights of each compound by subtracting
off-resonance data from the on-resonance data.  In this process,
we use the on-resonance spectra at slightly lower ($\sim$ 0.4 eV)
incident photon energy than the $M_5$ maximum peak of XAS, since
the spectra taken at the $M_5$ maximum peak position are often
found to be contaminated by incoherent Auger emissions\cite{Cho}.

\begin{figure}
\includegraphics[width=0.35\textwidth]{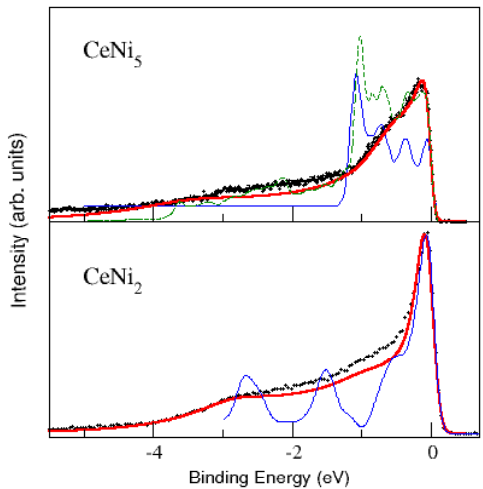}
\caption[] {Comparison between the experimental Ce $4f$ spectral
weights (dots) with band structure calculations (thin solid line)
and with GS fits (thick solid line) for CeNi$_x$ ($x$ = 2, 5). The
dashed line in the top panel is another band calculation by Harima
\cite{harima}.} \label{fig:5}
\end{figure}

The experimental bulk-sensitive $4f$ spectral weights of CeNi$_2$
and CeNi$_5$ thus obtained are shown as black dots in Figure 2.
Previously reported high resolution ($\Delta$E~$\sim$~50 meV) $4d
\rightarrow 4f$ resonant photoemission\cite{Purdie} observed the
$4f$-driven spectral feature less clear due to the dominant TM
$3d$ emission.  However, present work evidently shows the
$4f$-driven peak at the Fermi level in both Ce-Ni compounds. To
determine whether ``localized'' or ``itinerant'' picture is more
appropriate for these $4f$ spectral weights, we plot and compare
with both the Ce 4$f$ density-of-states (DOS) from linear
muffin-tin orbital (LMTO) calculations with local density
approximation (thin solid line)\cite{Lukas}, and the Gunnarsson
Sch\"onhammer (GS) calculation fit (thick solid line) based on the
Anderson single impurity model\cite{Gunn}. The one-electron $4f$
DOS were taken from the published band structure
calculations\cite{Lukas}, which were convoluted with Lorentzian
broadening in the form of $\alpha|\omega-\omega_0|$ and the
experimental resolution of 100 meV.  For the GS calculations the
lowest order $f^0$, $f^1$, $f^2$, and the second-order $f^0$
states are employed as basis states, and the spin-orbit splitting
of the 4$f$ level is included.  When GS calculation is executed,
it is known to be quite essential to employ realistic ${\rm
V}^2(\epsilon)$ in order to fully interpret experimental spectra
and fit the thermodynamic quantities \cite{Yang}. Hence we used
the $3d\rightarrow4f$ off-resonance spectrum for the valence band
shape to be hybridized with Ce 4$f$ state.

\begin{table}
\caption[] {The parameter values used for GS fitting and the
resulting $4f$-level occupancy number $n_f$ and the Kondo
temperature $T_{\rm K}$. Coulomb energy $U_{f\!f}$ is set to 6.0
eV.  $\chi_{m}(0)$'s are magnetic susceptibility at $T$=0
predicted from GS fitting and $\chi^{\ast}_{m}(0)$'s are
experimental values in units of $10^{-3}emu/mol$.}
\begin{center}
\begin{tabular}{l c r r r c c}
           &$\varepsilon_f$(eV)&$\Delta$(meV)&$T_K$(K)&$n_f$&$\chi_{m}(0)$&$\chi^{\ast}_{m}(0)$ \\ \hline
CeRh$_2$   &  1.30   &  95    & 1335  &  0.76  &  0.92   &  0.6\\
CeRh$_3$   &  1.20   &  110   & 1350  &  0.70  &  0.54   &  0.4\\
CeNi$_2$   &  1.13   &  89    &  570  &  0.78  &  1.13   &  0.9\\
CeNi$_5$   &  1.00   &  90    & 3300  &  0.69  &  0.62   &  0.7\\
\end{tabular}
\end{center}
\label{table:1}
\end{table}

From these comparisons, we can see that the GS calculations
provide quite good fits for the experimental Ce $4f$ spectral
weights for both CeNi$_2$ and CeNi$_5$. Their resulting parameter
values are presented in Table \ref{table:1} along with the Kondo
temperature $T_K$, the $4f$ electron occupation number $n_f$, the
magnetic susceptibilities at $T$~=~0, $\chi_{m}(0)$ deduced from
these parameter values and the experimentally measured
$\chi^{\ast}_{m}(0)$ \cite{susc}. We find that $T_K$ increases and
$n_f$ becomes smaller as the Ni content is increased, which is
consistent with the findings of XAS and other spectroscopic
investigations\cite{Fuggle83}, and can be attributed to the shift
of the Ni $3d$ valence band toward to the Fermi level at high Ni
concentration\cite{Purdie}. We also find that the $\chi_{m}(0)$
deduced from GS calculation gives very close values to measured
ones for both Ce-Ni compounds.

On the other hand, the band structure calculation gives rather
poor agreement with the experimental data for both compounds,
especially in CeNi$_5$. The calculation does not reproduce the
peak near the Fermi level properly, and the predicted strongest
peak around 1 eV from the Fermi level is absent in the
experimental data.  This discrepancy is not due to the particular
calculation method or misplacement of the Fermi level, since more
recent band calculation from other group (dot-dashed line in the
figure) also shows similar discrepancy\cite{harima}. For the cases
of CeNi$_2$, the band calculations reproduce the peak at the Fermi
level properly, but the features between 1-3 eV from the Fermi
level shows appreciable discrepancy. These disagreements of band
descriptions imply that the correlation between the $4f$ electrons
should not be neglected in these CeNi compounds, even in the
extremely $\alpha$-like compound CeNi$_5$.

\begin{figure}
\includegraphics[width=0.33\textwidth]{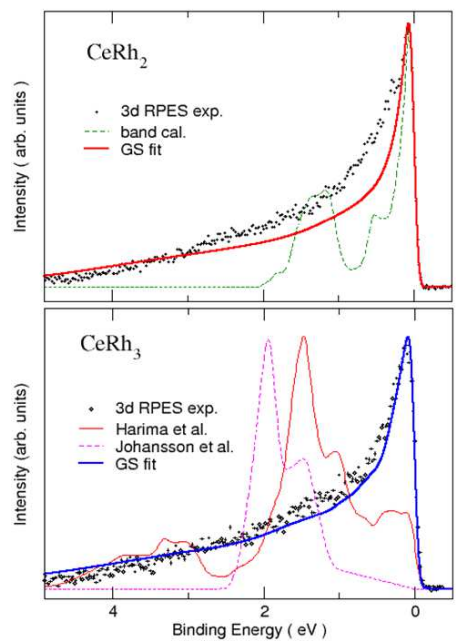}
\caption[] {The comparisons of experimentally extracted {\it
bulk-sensitive} $4f$ weights of CeRh$_2$ and CeRh$_3$ with the
one-electron band calculations and the GS fit.} \label{fig:6}
\end{figure}

The same phenomena happen in CeRh$_x$ (x=2,3) compounds, as can be
seen in Figure 3.   In this figure, the experimental
bulk-sensitive $4f$ spectral weights \cite{Expt} (dots) are
compared with the one-electron band structure calculation (thin
solid line) and the fit of the GS calculation (thick solid line).
In both CeRh$_2$ and CeRh$_3$ the experimental data show a large
peak near the Fermi level reminiscent of the Kondo tail, which can
be well described by the GS calculation.  However, the band
structure calculations done by two independent
groups\cite{Lukas,harima} completely miss this feature in
CeRh$_3$, and predict instead a strong peak near 2 eV from $E_{\rm
F}$, which is absent in the experimental data.   Even in the case
of CeRh$_2$, the band structure correctly predicts a peak near
$E_{\rm F}$, but again DOS below 2 eV from $E_{\rm F}$ shows
substantial disagreement.  On the other hand, GS calculations give
good general agreements in both Ce-Rh compounds, although some
discrepancy in intensity around 1 eV from $E_{\rm F}$ is seen in
CeRh$_2$.  Such disagreement seems to be commonly observed in
Laves-phase CeTM$_2$ compounds\cite{jnj}, and its origin is
unknown at present.  We list the parameter values of GS fitting
for CeRh$_2$ and CeRh$_3$ in Table \ref{table:1} along with their
$T_K$, $n_f$, $\chi_{m}(0)$ and $\chi^{\ast}_{m}(0)$. We find that
$\chi_{m}(0)$ is in good agreement with $\chi^{\ast}_{m}(0)$ in
both Ce-Rh compounds as in Ce-Ni case.

It may not be surprising that the band calculation does rather
poorly in describing photoemission spectra although it can
correctly predict volume anomaly and magnetic moment, since the
local density approximation is strictly valid only for the ground
state. However many physical properties such as specific heat
depend on the excitation spectrum, and the fact that itinerant
$4f$ band picture fails to describe the single-particle excitation
such as photoemission implies that it will not be able to
understand many physical properties of the system.  On the other
hand, Anderson impurity model gives consistent description of both
$4f$ spectral weights and thermodynamic properties with one set of
parameter values, as seen above in Figs. 2-3 and Table
\ref{table:1}. This shows that electron correlation effect is
important even for these extremely $\alpha$-like Ce compounds, and
that the localized $4f$ picture may be a better starting point
than the itinerant band picture to understand their physical
properties.

In summary, the {\it bulk-sensitive} Ce $4f$ spectral weights
obtained by $3d\rightarrow4f$ RPES of CeNi$_x$ (x=2, 5) and
CeRh$_x$ (x=2, 3) show remarkably enhanced the $f^1$ final state
peak near the Fermi level unlike those of $4d\rightarrow4f$ RPES.
The Gunnarsson Sch\"onhammer calculations based on the Anderson
single impurity model fit very well the experimental Ce 4$f$
spectral weights, and its resulting parameter values are
consistent with other spectroscopic and thermodynamic data.
However, one-electron band calculations fail to give plausible
descriptions.  This implies that the $4f$ states even in extremely
$\alpha$-like CeRh$_2$, CeRh$_3$, and CeNi$_5$ still retain
`localized' $4f$ character, at least for the excitation spectra.

This work is supported by the Korean Science and Engineering
Foundation through the Center for Strongly Correlated Materials
Research at Seoul National University. The experiments were
performed under the approval of the Japan Synchrotron Radiation
Research Institute.  We thank Prof. Harima for sending band
calculation results prior to publication.


\begin{references}
\bibitem{Expt}
Our 3d RPES data on CeRh$_3$ is somewhat different from the old
data\cite{Kaindl} in that the peak at the Fermi level is much
stronger and quite narrow.  This is probably due to the difference
in energy resolution and the incident photon energy.
\end{references}
\end{document}